\documentclass{eas}
\usepackage{graphicx}
\begin{document}

\title{Polycyclic Aromatic Hydrocarbons as Star Formation Rate Indicators} 
\author{Daniela Calzetti}\address{Dept. of Astronomy, University of Massachusetts, Amherst, MA 01003, USA; calzetti@astro.umass.edu}
%
%
\begin{abstract}
As images and spectra from ISO and Spitzer have provided increasingly higher--fidelity representations of 
the mid--infrared (MIR) and Polycyclic Aromatic Hydrocarbon (PAH) emission from galaxies and  galactic and extra--galactic regions, more systematic efforts have been devoted to establishing whether the emission in this wavelength region 
can be used as a reliable star formation rate indicator. This has also been in response to the extensive surveys of distant galaxies that have accumulated during the cold phase of the Spitzer Space Telescope. 
Results so far have been somewhat contradictory, reflecting 
the complex nature of the PAHs and of the mid--infrared--emitting dust in general. The two main problems faced when  
attempting to define a star formation rate indicator based on the mid--infrared emission from galaxies and star--forming regions 
are: (1) the strong dependence of the PAH emission on metallicity; (2) the heating of the PAH dust by evolved stellar populations 
unrelated to the current star formation. I review the status of the field, with a specific focus on these two problems, and will try 
to quantify the impact of each on calibrations of the mid--infrared emission as a star formation rate indicator. 
\end{abstract}
\runningtitle{D. Calzetti: PAHs as SFR Indicators}
\maketitle

\section{Introduction}
First enabled by the data of the Infrared Space Observatory (ISO, Kessler {\em et al} \cite{Kess1996}) and 
then expanded by the images and spectra of the Spitzer Space Telescope (Spitzer, Werner {\em 
et al.} \cite{Wern2004}), the mid--infrared, $\sim$3--40~$\mu$m, wavelength range has been at the 
center of many investigations seeking to define handy and easy--to--use Star Formation Rate (SFR) 
indicators,  especially in the context of extragalactic research.  

In the Spitzer (cold--phase) era, the MIR emission from galaxies had experienced renewed interest  
particularly in the high--redshift galaxy populations community (e.g., Daddi {\em et al.} \cite{Dadd2005,Dadd2007}, 
Papovich {\em et al.} \cite{Papo2006,Papo2007}, Yan {\em et al.} \cite{Yan2007}, Reddy {\em et al.} \cite{Redd2010}). 
The Spitzer Multiband Imaging 
Photometer  (MIPS, Rieke {\em et al.} \cite{Riek2004}) and the Spitzer InfraRed Spectrograph (IRS, Houck {\em et al.} 
\cite{Houc2004}) had proven particularly sensitive at detecting the MIR dust emission from galaxies in the redshift range 
z$\approx$1--3. Observations performed with the MIPS 24~$\mu$m and 70~$\mu$m bands yield the 
restframe $\sim$8~$\mu$m and 23~$\mu$m emission of a z=2 galaxy. An obvious question to ask is whether, 
and to what degree, the MIR emission traces the recent star formation in galaxies. 

The MIR region hosts the rich spectrum of the Polycyclic Aromatic Hydrocarbon 
(PAH) emission features in the wavelength range 3--20~$\mu$m (L\'eger \& Puget \cite{Lege1984}), also 
previously known as Unidentified Infrared Bands or Aromatic Features in Emission (e.g., Sellgren, Werner 
\& Dinerstein \cite{Sell1983}, Sellgren \cite{Sell1984}, Puget, L\'eger \& Boulanger \cite{Puge1985}), which 
are the subject of the present Conference.  Thus, a corollary to the question in the previous paragraph is how well the PAH 
emission traces recent star formation in galaxies. 

In this Review, I concentrate on results obtained from the investigation of nearby (closer than about 30--50~Mpc) 
galaxies, where the MIR emission can be typically resolved to sub--kpc scale at the Spitzer resolution in the 3--8~$\mu$m 
range: the 2$^{\prime\prime}$ FWHM corresponds to $\sim$0.5~kpc for a galaxy at 50~Mpc distance. In this regime, 
separating the contributing heating effects of various stellar populations become easier than in unresolved 
galaxies. Throughout this paper, I will refer to `8~$\mu$m emission' when indicating the stellar--continuum subtracted 
emission as detected by Spitzer in the IRAC 8~$\mu$m 
band (Fazio {\em et al.} \cite{Fazi2004}) or similar--wavelength bands on the ISO satellite. With the terminology `PAH emission' I will refer to the emission features (after subtraction of both the stellar and dust continuum) at/around 8~$\mu$m. 

An important assumption when using the infrared emission as a SFR indicator is that galaxies must contain sufficient dust 
that a significant fraction of the UV--optical light from recently formed stars is absorbed and re--emitted at longer 
wavelengths by the dust itself. For the `typical' galaxy in the Universe, about half of its energy budget is re--processed 
by dust in the infrared (e.g., Hauser \& Dwek \cite{Haus2001}, Dole {\em et al.} \cite{Dole2006}), but a very large scatter on 
this `mean' value exists from galaxy to galaxy. There tends to be a relation between star formation activity and dust 
opacity in galaxies, in the sense that more active galaxies also tend to be more opaque (e.g., Wang \& Heckman \cite
{Wang1996}, Heckman {\em et al.} \cite{Heck1998}, Calzetti \cite{Calz2001}, Sullivan {\em et al.} \cite{Sull2001}).  In general, infrared SFR calibrations will need to take into account that a fraction of the star formation will emerge from the galaxy 
unprocessed by dust, and this fraction will depend on a number of factors (metal content and star formation activity being 
two of those, and dust geometry being a third, harder--to--quantify, factor). 

\section{PAH Emission as a SFR Indicator}
The 8~$\mu$m emission represents about 5\%--20\% of the total infrared (TIR) emission from galaxies; of this, more than 
half, and typically 70\%,  can be attributed to PAH emission in metal--rich galaxies 
(Smith {\em et al.} \cite{Smit2007}, Dale {\em et al.} 
\cite{Dale2009}, Marble {\em et al.} \cite{Marb2010}). In this context, `metal--rich' refers to oxygen abundances 
larger than 12$+\log$(O/H)$\sim$8.1--8.2 (with the Sun's oxygen abundance being $\sim$8.7, Asplund {\em et al.} \cite{Aspl2009}). 
Although small, the fraction of TIR energy contained in the 7--8~$\mu$m wavelength region is not entirely negligible, and 
much effort has been devoted to investigating whether the emission in this region could be used as a reliable SFR indicator. 

Systematic efforts began with ISO (e.g., Roussel {\em et al.} \cite{Rous2001}, Boselli, Lequeux \& Gavazzi \cite{Bose2004}, Forster--Schreiber {\em et al.} \cite{Fors2004}, Peeters {\em et al.} 
\cite{Peet2004}), and continued with Spitzer, which enabled extending the analysis to fainter systems and to higher spatial detail  (e.g., Wu {\em et al.} \cite{Wu2005}, Calzetti {\em et al.} \cite{Calz2005,Calz2007}, Alonso--Herrero {\em et al} \cite{Alon2006}, Zhu {\em et al.} 
\cite{Zhu2008}, Kennicutt {\em et al.} \cite{Kenn2009}, Salim {\em et al.} \cite{Sali2009}, Lawton {\em et al.} \cite{Lawt2010}).  In a typical approach, the 8~$\mu$m (or similar wavelengths) luminosity or luminosity/area is plotted as a function of other known SFR indicators, to establish whether a correlation exists; an example is shown in Figure~1. In one case, the 8~$\mu$m emission has also been combined with tracers of the 
unattenuated star formation (e.g., H$\alpha$ line emission) to recover an `unbiased' SFR indicator (Kennicutt {\em et al.} \cite{Kenn2009}). 

\begin{figure}[ht]
 \includegraphics[width=9cm]{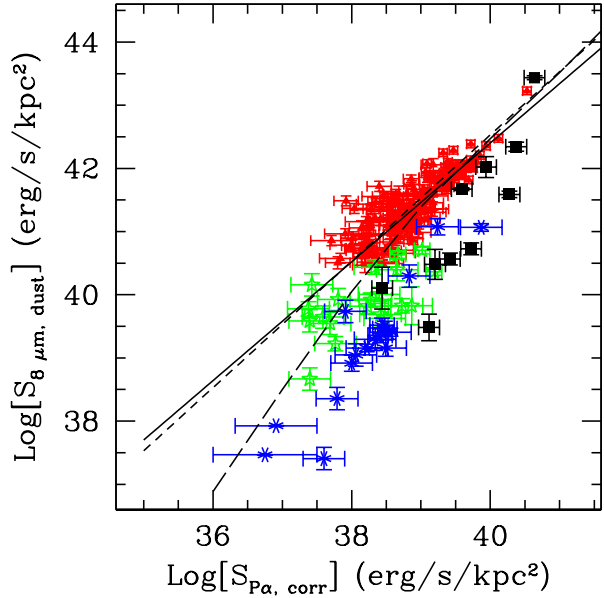}
 \caption{The surface luminosity density at 8~$\mu$m (stellar--continuum subtracted) as a function of the surface 
 luminosity density of the extinction--corrected hydrogen recombination emission line P$\alpha$ (1.8756~$\mu$m), for 
 HII~knots (red, green, and blue symbols) and low--metallicity starburst galaxies (black symbols). The HII-knots are typically 
 $\approx$0.5~kpc--size line--emitting regions from 33  nearby galaxies; there are a total of 220 regions in the plot. 
 The starburst galaxies are from the sample of  Engelbracht et al. (\cite{Enge2005}). The red symbols show regions hosted in galaxies with oxygen abundance 12$+$log(O/H)$>$8.35. Regions of lower 
 metal abundance are divided between those with oxygen abundance 8.00$<$12$+$log(O/H)$\le$8.35 (green symbols) and those with 12$+$log(O/H)$\le$8.00 (blue symbols). The continuous line shows the best linear fit (in the log--log plane)
 through the high metallicity data points (red symbols). The dash line shows the best linear fit with unity slope. A common result is for the best fit to yield a slope lower than unity. The dot-dash curve shows the locus of a model for which regions/galaxies suffer from decreasing dust attenuation (and, consequently,  decreasing dust emission) as their total luminosity decreases, a commonly observed trend in galaxies and star--forming 
 regions (see references in the last paragraph of the Introduction). Even when taking this effect into account, low metallicity regions and starburst galaxies display a depressed 8~$\mu$m emission, which has been shown to be due to deficiency of PAH emission. From Calzetti et al. (\cite{Calz2007}). }
\end{figure}

Most of these analyses recover a linear (in log--log space) relation  with slope of unity, or slightly less than 1, between the 8~$\mu$m emission and the reference SFR indicator. A roughly linear trend with unity slope would in general indicate that the 8~$\mu$m luminosity is an excellent SFR tracer; however, as we will see in the next two sections, the PAH emission (and, therefore, the 8~$\mu$m emission, of which the PAH emission represents typically more than half of the total) shows a strong dependence  on the metal content of the region or galaxy, and a less strong, but possibly significant, dependence on the nature of the heating stellar population. 

\section{The Dependence of PAHs on Metallicity}

Already known for more than a decade (e.g., Madden \cite{Madd2000}, and references therein), the dependence of the 
intensity of the PAH emission on the galaxy/environment metallicity has been quantified in detail by Spitzer data 
(e.g., Figure~1). This 
dependency is {\em on top} of the general decrease of the TIR intensity for decreasing metallicity, i.e. the decrease in 
overall dust content. Analyses of galaxy samples and of galaxy radial profiles covering a range of metallicity show an additional order--of--magnitude decrease in the 8~$\mu$m--to--TIR luminosity for  a factor $\sim$10 decrease in metallicity, with a transition at 12$+$log(O/H)$\approx$8.1 (Boselli, Lequeux \& Gavazzi \cite{Bose2004},  Madden {\em et al.} \cite{Madd2006}, Engelbracht {\em et al.} \cite{Enge2005}, Hogg {\em et al.} \cite{Hogg2005}, Galliano  {\em et al.} \cite{Gall2005,Gall2008}, Rosenberg  {\em et al.} \cite{Rose2006}, Wu {\em et al.} \cite{Wu2006}, Draine  {\em et al.} \cite{Drai2007}, Engelbracht  {\em et al.} \cite{Enge2008}, Gordon  {\em et al.} \cite{Gord2008}, Mu\~noz--Mateos  {\em et al.} \cite{Muno2009}, Marble  {\em et al.} \cite{Marb2010}). 

MIR spectroscopy with Spitzer has established that the observed low abundance of PAHs in low metallicity systems  is {\em not} due to the molecules being more highly ionized and/or dehydrogenated than in higher metallicity galaxies (Smith {\em et al.} \cite{Smit2007}), although there is controversy on whether they could be characterized by smaller or larger sizes (Hunt {\em et al.} \cite{Hunt2010}, Sandstrom {\em et al.} \cite{Sand2010}; see, also, Hunt et al. elsewhere in this volume and Sandstrom et al. elsewhere in this volume).

The nature of the correlation between the strength of the PAH emission features and the region's metallicity is still 
ground for debate. The two main scenarios that can account for the observed trend are: the `nature' scenario, according to 
which lower metallicity systems have a delayed PAH formation, and the `nurture' scenario, according to which PAHs are 
processed and/or destroyed in the harder radiation fields of low metallicity galaxies. 

Galliano, Dwek \& Chanial (\cite{Gall2008}, see, also, Galliano elsewhere in this volume) suggest that the PAH strength--metallicity correlation may be due to the delayed formation of the PAHs themselves, which are thought to form in the envelopes of carbon--rich AGB stars. In a young system, the first dust will emerge from supernovae (timescale$<$10~Myr), while AGB--produced dust will emerge at a later stage (timescale$\approx$1~Gyr). This scenario may be difficult to reconcile with the fact that low--metallicity systems in the local Universe contain stellar populations that are typically older than 2~Gyr (Tosi \cite{Tosi2009} and references therein), and have median birth parameters that are not drastically different from those of more metal rich galaxies (Lee {\em et al.} \cite{Lee2007}). However, low--mass systems are also thought to lose most of their metals during the sporadic events of star formation that characterize their typical star formation history (e.g., Romano, Tosi \& Matteucci \cite{Roma2006}). Clearly, self--consistent models for the metal enrichment of dwarf, low--metallicity  galaxies that also include their extended star formation histories are needed in order to accept or refute the `nature' scenario. 

Alternately, processing of PAHs by hard radiation fields has been proposed as a mechanism for the observed correlation between PAH luminosity and metallicity (Madden  {\em et al.} \cite{Madd2006}, Wu  {\em et al.}  \cite{Wu2006}, Bendo  {\em et al.}  \cite{Bend2006}, Smith {\em et al.} \cite{Smit2007}, Gordon {\em et al.} \cite{Gord2008}, Engelbracht {\em et al.} \cite{Enge2008}), since low--metallicity environments are generally characterized by harder radiation fields than high--metallicity ones (e.g. Hunt {\em et al.} \cite{Hunt2010}). Additionally, efficient destruction of PAHs by supernova shocks may also contribute to the observed deficiency (O'Halloran, Satyapal \& Dudik \cite{OHal2006}). These suggestions agree with the observation that PAHs are present in the PDRs surrounding HII regions, but are absent (likely destroyed) within the HII regions (Cesarsky {\em et al.} \cite{Cesa1996}, Helou {\em et al.} \cite{Helo2004}, Bendo {\em et al.} \cite{Bend2006}, Rela\~no \& Kennicutt \cite{Rela2009}). A recent study of the SMC finds a high fraction of PAHs within molecular clouds (Sandstrom {\em et al.} \cite{Sand2010}), thus complicating the interpretation of the formation/destruction mechanisms for these molecules. Both production and processing may ultimately be driving the observed  trend (Wu {\em et al.} \cite{Wu2006}, Engelbracht {\em et al.} \cite{Enge2008}, Marble {\em et al.} \cite{Marb2010}).

\section{The Heating Populations of PAHs}

Outside of HII regions and other strongly ionizing environments, PAHs tend to be ubiquitous. These large molecules are transiently heated by single UV and optical photons, and, therefore, they can be heated by the radiation from the mix of stellar populations that contribute to the general interstellar radiation field, unrelated to the current star formation in a galaxy  (Haas, Klaas \& Bianchi \cite{Haas2002}, Boselli, Lequeux \& Gavazzi \cite{Bose2004}, Peeters, Spoons \& Tielens \cite{Peet2004}, Mattioda, Allamandola \& Hudgins \cite{Matt2005}, Calzetti {\em et al.} \cite{Calz2007}, Draine \& Li \cite{Drai2007a}, Bendo {\em et al.} \cite{Bend2008}). 

Bendo {\em et al.} (\cite{Bend2008}) show that the 8~$\mu$m emission from 
galaxies is better correlated with the 160~$\mu$m dust emission than with the 24~$\mu$m emission, on spatial scales $\approx$2~kpc. Haas Klaas \& Bianchi (\cite{Haas2002}) also find that the ISO 7.7~$\mu$m emission is correlated with the 
850~$\mu$m emission from galaxies. Boselli, Lequeux \& Gavazzi (\cite{Bose2004}) show that the stellar populations responsible for the heating of the MIR--emitting dust are more similar to those responsible for heating the large grains 
emitting in the far--infrared, rather than to those responsible for the ionized gas emission in galaxies. This result is similar to that obtained by Peeters, Spoons \& Tielens (\cite{Peet2004}) for a combined Galactic and extragalactic sample of PAH--emitting sources. All the evidence points to a close relation between the MIR/PAH 
emission and the cold dust heated by the general (non--star--forming) stellar population.

The presence of  heating by evolved stellar populations thus produces a ill--quantified 
contribution to the total MIR luminosity of a galaxy, affecting the calibration of any SFR indicator using that wavelength range (e.g., Alonso--Herrero {\em et al.} \cite{Alon2006}). We still do not have a clear understanding of dependencies on galaxy morphology, stellar population mix, star formation rate, star formation intensity, etc. 

A recent analysis of the nearby galaxy NGC0628, an almost face--on SAc at a distance of about 7.3~Mpc, shows that its 8~$\mu$m emission contains about 20\%--30\%  contribution from a diffuse component unrelated to sites of current star formation (Crocker {\em et al.} \cite{Croc2010}). The 20\%--30\% range reflects uncertainties in the adopted extinction and [NII] contamination corrections for the H$\alpha$ emission, used here to trace sites of recent star formation. We can also obtain a rough idea of the amount of diffuse emission in the MIR bands by converting the 8~$\mu$m mean  luminosity 
density, $\sim$1.2$\times$10$^7$~L$_{\odot}$~Mpc$^{-3}$, within the local 10~Mpc (including emission from 
both PAHs and dust continuum, see Marble {\em et al.} \cite{Marb2010}) to a volume SFR density, using 
metallicity--dependent linear calibrations to SFR(8~$\mu$m) from the data of Calzetti {\em et al.} (\cite{Calz2007}). The result, $\rho_{SFR}$(8~$\mu$m)$\sim$0.019~M$_{\odot}$~yr$^{-1}$~Mpc$^{-3}$, is roughly 30\%--60\% higher than the commonly accepted values for the SFR density in the local Volume (see references in Hopkins \& Beacom \cite{Hopk2006}).

While the above suggests that the contribution to the 8~$\mu$m emission from dust heated by the diffuse stellar populations is not large (less than a factor $\approx$1.5--2), when galaxies are considered as a whole or as {\em populations}, we should recall that we don't have a handle on galaxy--to--galaxy variations. Even worse, the diffuse population heating could become a prevalent contribution to the 8~$\mu$m emission in some galactic regions, thus potentially affecting any investigation of spatially resolved features within galaxies. 

\section{Summary and Conclusions}

Over the past decade, the accumulation of both spectroscopy and high--angular--resolution imaging of the MIR emission 
from galaxies has paved the road for an accurate investigation of the MIR/PAH luminosity as a SFR tracer. Data have shown that there is generally a linear (with slope of 1 or slightly less than 1 in a log--log plane) correlation between the stellar--continuum--subtracted MIR luminosity and the SFR, for {\em high metallicity} systems, i.e., for galaxies and regions that are about 1/5--1/3  solar or higher in oxygen abundance. While a linear correlation would generally indicate that the MIR emission can be considered a reliable SFR indicator, there are at least two caveats to keep in mind: (1) the PAH emission has a strong dependence on metallicity; (2) the MIR--emitting dust can be heated by evolved stellar populations unrelated to the current star formation. 

The PAH dependence on metallicity is well established and quantified, and shows an order--of--magnitude deficiency in PAH/TIR emission for a decrease in oxygen abundance by about a factor 10. The nature of this dependency is still debated, and could be due to delayed production of the PAHs in low metallicity systems or to processing/destruction mechanisms in the harder radiation fields of low metallicity galaxies, or a combination of both.

The heating of PAHs by evolved stellar populations is also well established, but less well quantified. It is likely to have a smaller impact on any SFR(PAH) or SFR(MIR) calibration than the metallicity dependence, probably at the level of less than a factor 2. However, potential variations as a function of galaxy morphology, stellar population mix, star formation rate, star formation intensity, etc., and the impact of the diffuse population heating as a function of location for sub--galactic scale analyses have not been quantified yet. This is a virtually unexplored realm with a possibility for major implications as many investigations move from the `global populations' approach to the `sub--kpc regions' approach in the study of star formation in galaxies.


\end{document}